\PassOptionsToPackage{dvipsnames}{xcolor}
\documentclass[twocolumn,twocolappendix]{aastex702}

\newcommand{\kms}{\ensuremath{\mathrm{km~s}^{-1}}}
\newcommand{\mbh}{\ensuremath{M_\mathrm{BH}}}
\newcommand{\mlgen}{\ensuremath{M^*/L}}
\newcommand{\ml}{\ensuremath{M^*/L}}
\newcommand{\msun}{\ensuremath{M_{\odot}}}
\newcommand{\lsun}{\ensuremath{L_{\odot}}}
\newcommand{\rsoi}{\ensuremath{r_\mathrm{SOI}}}

\begin{document}

\title{The M87 Black Hole Mass from JWST and Keck Stellar Kinematics Spanning 10~pc to 10~kpc}

\author
[0000-0002-7703-7077]
{Emily R. Liepold}
\affiliation{Department of Astronomy, University of California, Berkeley, California 94720, USA}
\email{emilyliepold@berkeley.edu}

\author
[0000-0002-4430-102X]
{Chung-Pei Ma}
\affiliation{Department of Astronomy, University of California, Berkeley, California 94720, USA}
\affiliation{Department of Physics, University of California, Berkeley, California 94720, USA}
\email{cpma@berkeley.edu}

\author
[0009-0008-8003-1048]
{Refa M.~Al-Amri}
\affiliation{George P.~and Cynthia Woods Mitchell Institute for Fundamental Physics and Astronomy, and Department of Physics and Astronomy, Texas A\&M University, College Station, TX 77843, USA}
\email{refaalamri@tamu.edu}

\author
[0000-0002-1881-5908]
{Jonelle L. Walsh}
\affiliation{George P.~and Cynthia Woods Mitchell Institute for Fundamental Physics and Astronomy, and Department of Physics and Astronomy, Texas A\&M University, College Station, TX 77843, USA}
\email{walsh@tamu.edu}

\begin{abstract}

We combine JWST NIRSpec and Keck KCWI observations to construct a contiguous 2D map of M87's stellar kinematics spanning three decades in projected radius, from ${\sim}10$~pc to ${\sim}10$~kpc, and measure the mass of its central supermassive black hole.
The innermost stellar kinematic measurements probe projected radii as small as ${\sim}3$\% of the black hole's sphere of influence radius, or approximately 30,000 Schwarzschild radii.
The NIRSpec spectra within $R < 0\farcs45$ are contaminated by quasiperiodic ``wiggle'' artifacts.
We develop \texttt{hyphen}, which uses detector spectral traces to predict how the wiggle frequency varies with wavelength, then fits and subtracts the wiggle pattern from each spectrum in each exposure before combining the data.
The cleaned spectra reveal a pronounced central rise in stellar velocity dispersion and $h_4$.
Triaxial Schwarzschild stellar orbit modeling with the TriOS code, fitting ${\sim}5,000$ kinematic constraints across 618 spatial bins from JWST and Keck, yields a black hole mass of $\mbh=(5.38 \pm 0.13\pm 0.22)\times10^9\,\msun$.
This result is consistent with our prior KCWI-only determination but with the statistical uncertainty reduced by a factor of 2--3, now subdominant to the systematic uncertainty in the distance ($D$) to M87.
Our inferred mass-to-distance ratio is $\theta_g=G\mbh/(c^2 D) = 3.16 \pm 0.08\, \mu$as, whose 90\% credible interval overlaps with that obtained from the multi-epoch Event Horizon Telescope (EHT) analysis, providing a robust and independent prior for future EHT studies.

\end{abstract}


\section{Introduction}
\label{sec.intro}

The supermassive black hole (SMBH) at the center of the giant elliptical galaxy M87 has anchored black hole research for half a century.
At a distance of only 16.8 Mpc, the M87 SMBH has one of the largest gravitational spheres of influence (SOI) of any SMBH on the sky (about $5\arcsec$), making M87 an excellent target for resolving stellar motions inside the region in which the SMBH determines the stellar kinematics.

\citet{sargentetal1978} observed a rising stellar velocity dispersion profile towards the center of M87 and interpreted it as dynamical evidence for a central mass of several billion solar masses.
\citet{binneymamon1982} argued, however, that spherical Jeans models without a central SMBH could fit the same data {\it if} the velocity dispersion is allowed to be anisotropic in the radial vs. tangential directions.
Subsequently, \citet{dresslerrichstone1990} obtained improved stellar spectra from the 2.5-m du Pont Telescope but reported an upper limit on the black hole mass (\mbh) of ${\sim} 4\times 10^9 \msun$ from spherical orbit modeling.
\citet{vandermarel1994} applied spherical anisotropic Jeans models to stellar kinematic data from the 4.2-m William Herschel Telescope, again finding an upper limit of ${\sim} 5\times 10^9 \msun$.
\citet{magorrianetal1998} applied two-integral axisymmetric Jeans models to the same kinematics and obtained $\mbh=(3.96\pm0.27)\times 10^9 \msun$.\footnote{All M87 SMBH masses quoted in this paper are scaled to our adopted distance of $16.8\pm 0.7$ Mpc.}
\citet{simonetal2024} explored axisymmetric anisotropic Jeans models of various sets of integral-field spectroscopic data and found a range of $\mbh = (7.7 - 11.1)\times 10^9 \msun$ depending on which dataset is used and whether dark matter is included in the models.
\citet{gebhardtetal2011} applied axisymmetric orbit modeling to integral-field spectroscopic data from Gemini NIFS, obtaining $\mbh = (6.14_{-0.62}^{+1.07}) \times 10^9\msun$ (after distance adjustment by \citealt{EHT_M87_2019F}).

To reduce modeling complexity and computational cost, all of these works assumed that the gravitational potential of M87 was either spherical or axisymmetric.
Massive elliptical galaxies such as M87, however, are found to be predominantly slowly rotating or non-rotating (e.g., \citealt{vealeetal2017a, vealeetal2017}).
And when rotation is detected, the kinematic axis connecting the regions of maximal receding and approaching velocities is often misaligned with the photometric major axis, a feature that cannot be explained by an axisymmetric stellar halo (e.g., \citealt{eneetal2018}).
Any spherical or axisymmetric dynamical model would therefore be unable to reproduce these kinematic misalignments in the data.

A prominent kinematic misalignment was indeed observed for M87 \citep{Emsellemetal2014, liepoldetal2023}.
Motivated by these findings, we removed the axisymmetric assumption for the first time and allowed M87's stellar halo to be triaxial \citep{liepoldetal2023}. 
In that work, we measured M87's stellar kinematics using 62 pointings of Keck KCWI observations spanning a $250\arcsec{\times}300\arcsec$ (20 kpc${\times}$24 kpc) field of view, extending from inside the SMBH's SOI to far into the stellar halo where the potential is dominated by dark matter.
By fitting the ${\sim}$4,000 kinematic constraints to triaxial Schwarzschild orbit models generated by the TriOS code \citep{quennevilleetal2022}, we obtained the first determination of the triaxial shape of M87's stellar halo and measurements of its \mbh, stellar mass-to-light ratio, and dark matter content.
That work found $\mbh = (5.37^{+0.37}_{-0.25}\pm 0.22)\times 10^9\msun$, where the first uncertainty is statistical and the second is systematic due to the uncertainty in the distance to M87.

A parallel literature on M87 SMBH mass measurements using kinematics of ionized gas has consistently yielded a smaller value in the range of $\mbh=(2.5-4)\times10^9\msun$ \citep{vandermarel1994, harmsetal1994, macchettoetal1997, walshetal2013}.
These measurements are complicated by the rich structures present in the ionized gas in the central 100 pc \citep{osornoetal2023}.

The Event Horizon Telescope (EHT) has imaged a ring of emission around the M87 SMBH, providing an independent constraint on its mass. 
Their first mass determination based on the 2017 data gave $\mbh=(6.5\pm 0.2 \pm 0.7)\times10^9\msun$ \citep{EHT_M87_2019F}, where the second uncertainty was due to various systematics including distance uncertainties.
A multi-epoch analysis combining the 2017 and 2018 data revised the mass to $\mbh = (6.16_{-0.57}^{+0.71})\times 10^9\msun$ \citep{EHT_M87_2025A}, where the reported error included statistical and systematic uncertainties but not the distance uncertainty (which is subdominant to their combined uncertainty).

In this Letter, we combine the Keck KCWI data from \citet{liepoldetal2023} with JWST NIRSpec data from \citet{al-amrietal2026} and extract ${\sim}5000$ stellar kinematic measurements spanning a radial range of ${\sim}10$ pc to 10 kpc of M87.
We use these kinematics as constraints in a new suite of triaxial Schwarzschild orbit models computed with the TriOS code to determine the black hole mass, stellar mass-to-light ratio, dark matter fraction, and 3D shape of the stellar halo of M87.
The new JWST measurements quadruple the total number of kinematic constraints within the SMBH's SOI compared to the Keck data alone, probing nearly an order of magnitude closer to the SMBH, down to 3\% of the SOI radius and only 30,000 times the Schwarzschild radius of M87's SMBH.

Sections~\ref{sec.obs}~and~\ref{sec.stellar_kinematics} describe our observations and stellar kinematic measurements.
Section~\ref{sec.modeling} details our triaxial orbit models and a new parameter inference method used to improve the efficiency of searching the 6D parameter space.
Section~\ref{sec.properties} discusses our updated measurements of M87's mass and shape parameters.
In Appendix~\ref{appendix:hyphen}, we describe our code \texttt{hyphen} designed to remove the prominent quasiperiodic artifacts in JWST NIRSpec spectra near M87's central point source.

In this paper, we adopt a distance to M87 of $16.8\pm0.7$ Mpc compiled in \citet{EHT_M87_2019F}.
An angular scale of 1\arcsec\ at this distance corresponds to a physical scale of $81.1\pm 3.3$~pc.

\section{Integral-Field Spectroscopic Observations of M87}
\label{sec.obs}

\subsection{Wide-field $250\arcsec{\times300}\arcsec$ Coverage with Keck KCWI}
\label{sec.obs.kcwi_data}

During four observing runs from 2020 to 2022, we used the Keck Cosmic Web Imager (KCWI; \citealt{Morrisseyetal2018}) to obtain integral-field spectra of M87 over a contiguous $250\arcsec{\times}300\arcsec$ (20 kpc ${\times}$ 24 kpc) field of view centered on the galaxy's nucleus. We constructed a mosaic from 62 pointings, with a total on-source exposure time of 13 hr. The observations used the large-slicer configuration with the BL grating to obtain broad spatial ($33\arcsec{\times}20\arcsec$ per pointing) and spectral (3500\text{--}5600~\AA) coverage.  With $1\farcs4 {\times} 0\farcs3$ spaxels and 0\farcs63 seeing, these observations allowed for stellar kinematic measurements down to a radius of ${\sim} 0\farcs8$ from M87's center. The KCWI spectra contain numerous optical absorption features, including the Ca~H and K lines, the G-band, Mgb, and several iron lines. The KCWI stellar kinematic measurements and orbit modeling results are presented in \citet{liepoldetal2023} and Section~\ref{sec.stellar_kinematics}.

\subsection{Central $3\arcsec{\times} 3\arcsec$ with JWST NIRSpec}
\label{sec.obs.jw_data}

We obtained JWST NIRSpec observations (Cycle 1 GO-2228) of the central $3\arcsec\times3\arcsec$ (240 pc $\times$ 240 pc) region of M87. The observations used the G235H grating with the F170LP filter, covering $1.66$--$3.17~\micron$ with $R \approx 2700$. The total on-source exposure time was 6.6 hr across 16 dithered exposures. The primary stellar absorption features in this window are the $K$-band CO bandheads near $2.3~\micron$, which we use to measure the stellar kinematics. The kinematic fitting strategy and results are presented in \citet{al-amrietal2026} and Section~\ref{sec.stellar_kinematics}.

\subsubsection{Outside the Central $R = 0\farcs45$ Region}
\label{sec.obs.outer_jw_data}

For the NIRSpec spaxels located outside a radius of $0\farcs45$ from the nucleus, we use the datacube reduced in \citet{al-amrietal2026}. A prominent ``wiggle'' artifact severely contaminates the spectra \citep[e.g.][]{lawetal2023} within $R{\sim}0\farcs45$, preventing robust stellar kinematic measurements without further processing. This region was therefore excluded from the analysis in \citet{al-amrietal2026}. For these G235H/F170LP observations, the PSF has an azimuthally averaged FWHM of 0\farcs113, with a characteristic snowflake pattern in its wings. The resulting spectra had typical SNRs of ${\sim}150$ per spectral pixel in individual $0\farcs05$ spaxels across much of the field. To perform orbit modeling, we rebin these spectra and remeasure the stellar kinematics as described below (Section~\ref{sec.stellar_kinematics}).

\subsubsection{Within $R = 0\farcs45$}
\label{sec.obs.inner_jw_data}

To extract useful science from the NIRSpec spaxels within $R=0\farcs45$ of M87's central AGN, we have developed an algorithm, which we name \texttt{hyphen}, to remove the spectral artifacts. Details of this method are described in Appendix~\ref{appendix:hyphen}. Here we provide a brief description.

In the JWST NIRSpec IFU and MSA modes, the PSF is undersampled, producing a Moir\'e-like pattern on the NIRSpec detectors. Once the 2D data are rectified into a 3D datacube, the spectra near a bright point source may contain a prominent quasiperiodic ``wiggle'' artifact associated with that Moir\'e pattern. This undersampling causes the distribution of flux across the detector pixels to depend strongly on the location of the point source relative to the pixel grid. If the point source lands near the center of a pixel, a larger fraction of its light falls within that pixel. If instead the point source lands near the edge, its light is split between pixels and the peak pixel flux is reduced \citep{lawetal2023}.

On the NIRSpec detector, the point-source light is dispersed along a trace that is roughly aligned with the detector $x$-axis. Because the trace is not parallel to that axis, it repeatedly crosses from one detector pixel to the next in the $y$ direction as the wavelength changes. This repeated crossing results in quasiperiodic variation in the pixel flux, which appears as a ``wiggle'' in the spectrum. The shapes of the spectral traces are shown in Figure~\ref{fig:a}, and the relation to the wiggle pattern is shown in Figure~\ref{fig:f}.

In \texttt{hyphen}, we use the World Coordinate System (WCS) transformations between the 2D detector images and the 3D rectified datacubes from each exposure to identify the sinusoidal basis functions associated with the wiggle pattern. This approach fits the repeating wiggle pattern while reducing the risk of fitting physical features that qualitatively resemble the artifact, such as the CO bandheads. 

A few other techniques have been developed to remove the NIRSpec artifacts. \texttt{WICKED} \citep{dumont_wiggle_2025} requires many sinusoidal fits over narrow spectral windows in spectra from the combined datacube to capture the varying frequency, while \texttt{RACCOON} \citep{shajib_textttraccoon_2025} simultaneously fits the wiggle shape and its frequency variation across the full spectrum. \texttt{ATM} \citep{lawclarke2026} relies on sampling different pixel phases and is less well constrained in parts of the spectrum where the wiggle frequency is low. \texttt{hyphen} instead uses the WCS transformations to constrain the frequency variation even where few wiggle cycles are present, and only uses the spectra to determine the shape of the repeating wiggle pattern. For our M87 observations, \texttt{hyphen} removes the vast majority of wiggle artifacts and leaves smaller residuals than \texttt{ATM}. Example spectra from the central region before and after our wiggle correction are shown in Figure~\ref{fig:b}.

Because \texttt{ATM} was introduced after the publication of \citet{al-amrietal2026}, we re-reduce the data presented there using the JWST SDP version \texttt{2025\_4a}, CRDS version \texttt{13.1.14}, context file \texttt{jwst\_1535.pmap}, and the JWST Science Calibration Pipeline version 2.0.0 so that a direct comparison between \texttt{ATM} and \texttt{hyphen} can be made. Bad pixels in each 2D detector image are manually flagged and masked at the end of stage 2 of the pipeline. Each detector image is then cast onto a common 3D grid with spacing 0\farcs05$\times$0\farcs05$\times$3.96~\AA\ and stored as a separate datacube. We apply \texttt{hyphen} to remove wiggle artifacts from each exposure's datacube and then merge the corrected cubes for spectral fitting.

We visually inspected the central spectra in the $\rm{CO}$ bandhead region around $2.3~\micron$ and separated the spaxels into three categories: AGN-like, ETG-like, and intermediate. The spectra within $R \lesssim 0\farcs15$ are dominated by the AGN continuum, with no obvious CO bandheads present and a positive spectral index. The spectra at $R \gtrsim 0\farcs21$ are consistent with a typical ETG CO bandhead region, including a negative spectral index and rounded bandheads. Between $0\farcs15$ and $0\farcs21$, many spaxels have intermediate features, including anomalous CO bandhead shapes or a positive or varying spectral index. Of the 250 spaxels with $R < 0\farcs45$, 28 are AGN-like, 27 are intermediate, and 195 are ETG-like. For the stellar kinematic measurements and orbit modeling described in the next sections, we use only the ETG-like spectra.

\section{Stellar kinematics from 10 pc to 10 kpc}
\label{sec.stellar_kinematics}

Stellar kinematics extracted from the Keck KCWI data are presented in our prior work \citep{liepoldetal2023}. Briefly, we co-added spectra from individual spaxels to produce 461 spatial bins, achieving signal-to-noise ratios (SNR) per spectral pixel of ${\sim}150$ in the central regions and ${\sim}100$ in the outer regions.\footnote{The SNRs in this paper are estimated using the $\beta\sigma$ procedure \citep{czeslaetal2018} using jump parameter $j=3$ and approximation order $N=3$.} We then measured the Gauss-Hermite moments of the stellar line-of-sight velocity distributions (LOSVDs) using pPXF \citep{cappellari_full_2023} and stellar templates from the MILES library \citep{sanchez-blazquezetal2006,falcon-barrosoetal2011}. The first eight moments were used as constraints in the triaxial orbit models presented in \citet{liepoldetal2023}.

For the JWST NIRSpec data, we use slightly different binning schemes for the regions inside and outside $R = 0\farcs45$. Outside $R = 0\farcs45$, we co-add the 3552 individual spectra using a Voronoi binning scheme \citep{CappellariCopin2003} to produce 133 spatial bins with a mean SNR of $308$ per spectral pixel. This binning choice roughly balances the number of kinematic constraints per logarithmic interval in radius between the NIRSpec data and the inner KCWI data: approximately 350 bins per dex in $R$ for both datasets. 

Inside $R = 0\farcs45$, we exclude the 28 AGN-like and 27 intermediate-type spaxels (see Sec.~\ref{sec.obs.inner_jw_data}) and bin the remaining 195 ETG-like spaxels on a regular polar grid.  We use three radial intervals, $R \leq 0\farcs3$, $0\farcs3 < R \leq 0\farcs38$, and $0\farcs38 < R \leq 0\farcs45$, and divide each annulus further into eight $45^\circ$ equiangular sectors. The sectors are defined relative to the photometric position angle, beginning with $-25^\circ < \mathrm{PA} < 20^\circ$. This arrangement produces 24 bins containing comparable numbers of spaxels (between 6 and 10 spaxels per bin) and a mean SNR of $340$.

\begin{figure}
  \centering
  \includegraphics[width=\columnwidth]{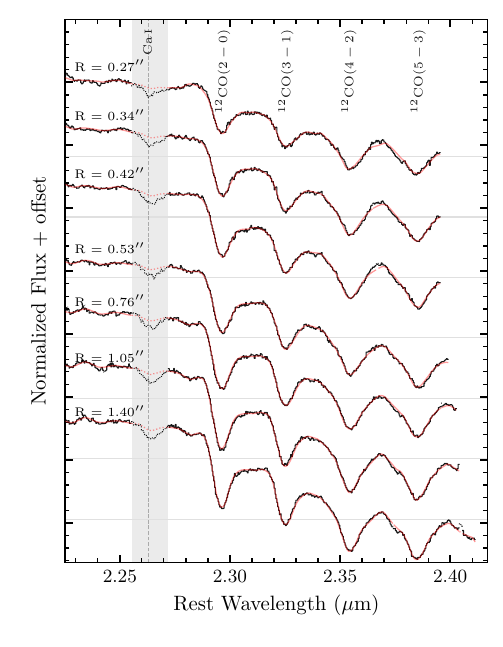}
  \caption{
  	Representative NIRSpec spectra (black) for seven of the 157 JWST bins at increasing radii (from top down) used to constrain the triaxial orbit models of M87. 
  	The pPXF fits to the spectra are shown in red. The PHOENIX stellar templates systematically misfit the Ca~I absorption feature at $2.26~\micron$, which we mask out (shaded).
  	The NIRSpec detector gap spans different wavelengths in different spatial bins, so the fitted spectral range varies between bins.
  }
\label{fig:cca}
\end{figure}

We use the same pPXF fitting configuration as in \citet{al-amrietal2026} to measure the stellar kinematics for each of the binned JWST spectra. The spectra are fit over $2.244-2.3632~\micron$ using 90 stellar templates generated with PHOENIX \citep{husser_new_2013}. The stellar LOSVDs are represented by 12 Gauss-Hermite moments, and the stellar continuum is modeled as an order-3 multiplicative polynomial. We determine the uncertainties on the kinematic moments using a Monte Carlo procedure in which each best-fitting pPXF model is perturbed with 1000 realizations of Gaussian noise having an RMS equal to that of the best-fitting residuals. 

Seven representative binned JWST spectra from $R=0\farcs27$ to $R=1\farcs40$ are shown in Figure~\ref{fig:cca}. The deep CO bandheads in each spectrum (black curves) are well matched by the stellar-template spectrum broadened by the best-fit LOSVD (red curves).

\begin{figure*}
  \centering
  \includegraphics[width=\textwidth]{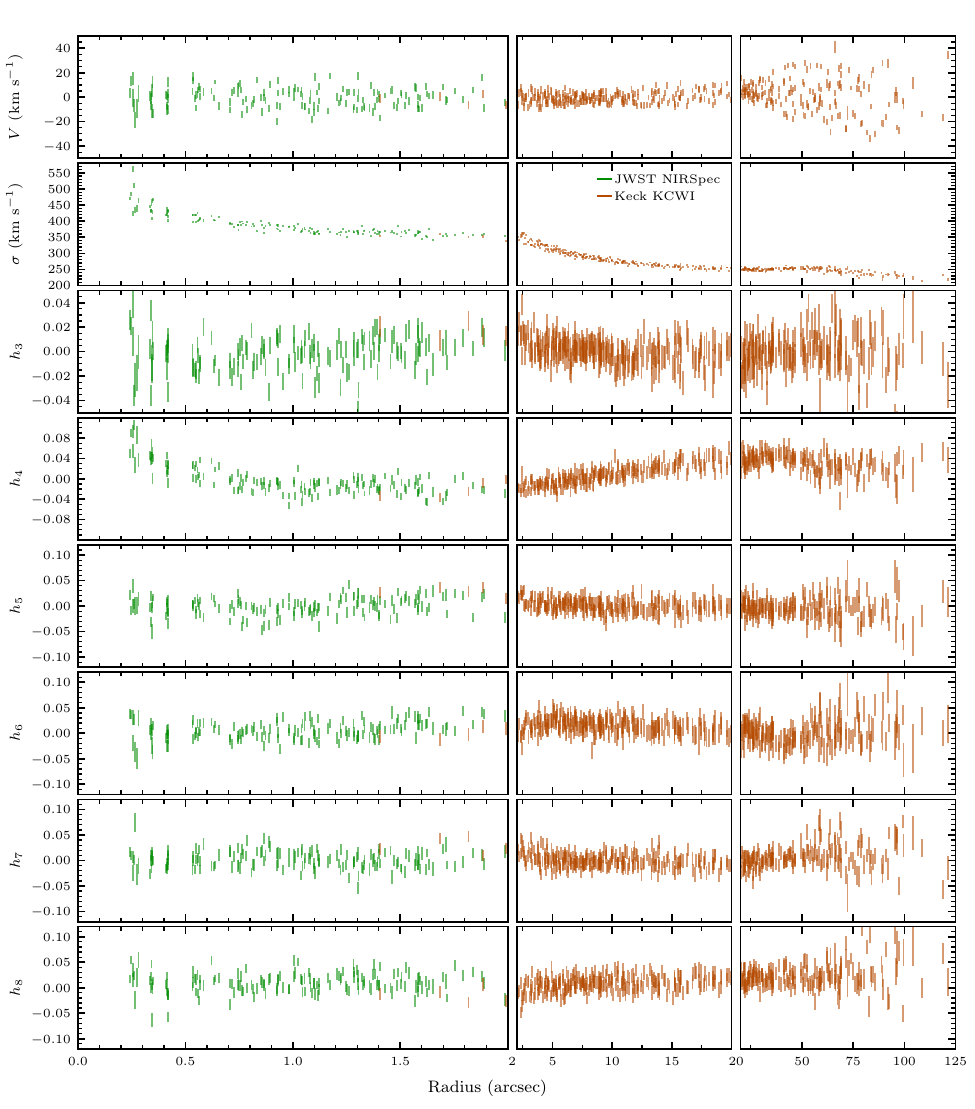}
  \caption{
	Radial profiles of the lowest eight Gauss-Hermite moments of the stellar LOSVDs measured from high-SNR spectra in 157 JWST bins (green) and 461 Keck bins (brown).
    To avoid crowding, the horizontal axis uses three different linear scales for the JWST kinematics
    ($R<2\arcsec$), inner KCWI kinematics ($R=2\arcsec-20\arcsec$) and outer KCWI kinematics ($R>20\arcsec$). The innermost five KCWI points lie within 2\arcsec and agree well with the JWST data.
  }
\label{fig:g}
\end{figure*}

The radial profiles of the first eight Gauss-Hermite kinematic moments measured from each of the 461 Keck and 157 JWST bins are shown in Figure~\ref{fig:g}. These 4944 measurements span nearly 3 decades in radii, which we use as kinematic constraints in the orbit models discussed in the next section. The JWST kinematic moments (green symbols) connect to the Keck moments (brown symbols) smoothly at $R{\sim}2\arcsec$ without systematic offsets. For the innermost five Keck bins (at $R{\sim}1\arcsec\text{--}2\arcsec$) that overlap with the JWST bins, the stellar kinematics are also remarkably consistent. This close match in the KCWI and NIRSpec results indicates that our measurements are robust to differences in telescopes, spectrographs, spectral wavelength range, and stellar template libraries used to determine the kinematics. 

\begin{figure}
  \centering
  \includegraphics[width=\columnwidth]{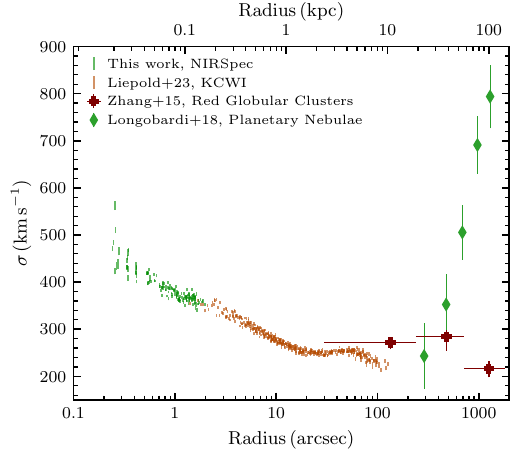}
  \caption{
    Profile of the stellar velocity dispersion of M87 over 4 decades in radius (on a logarithmic scale).
    The JWST NIRSpec data points (green) connect smoothly to the innermost Keck KCWI data points (brown) at $R{\sim}$1\arcsec--2\arcsec\ and extend the rising profile seen in the Keck data inward, reaching $\sigma{\sim}500$ \kms\ at $R{\sim}0\farcs2$.
    The Keck data in turn are consistent with $\sigma$ inferred from globular clusters and planetary nebulae at $R{\sim}$100\arcsec--300\arcsec\ \citep{zhangetal2015, Longobardietal2018}. 
  }
\label{fig:sigma}
\end{figure}

The velocity dispersion, $\sigma$, of the Keck data in Figure~\ref{fig:g} displays a prominent rise towards the nucleus of M87, from $\sigma{\sim}250$ \kms\ at $R{\sim}20\arcsec$ to $\sigma{\sim}350$ \kms\ at $R{\sim}2\arcsec$. The JWST data show that this rise continues smoothly all the way to $R{\sim}0.2\arcsec$, with $\sigma$ reaching ${\sim}500$ \kms. In Figure~\ref{fig:sigma}, we plot the profile of M87's stellar $\sigma$ over 4 decades in radius (on a logarithmic radial scale). In the outer part of M87's stellar halo, red globular clusters (red symbols) appear to have velocity dispersions similar to those of the stars \citep{zhangetal2015}. ``Intracluster'' planetary nebulae, on the other hand, have been reported to have a steeply rising $\sigma$ profile (green symbols; \citealt{Longobardietal2018}).

\section{Triaxial Orbit Modeling of M87}
\label{sec.modeling}

\subsection{The TriOS Code}
\label{sec.modeling.trios}

We perform triaxial orbit modeling with the \texttt{TriOS} code \citep{quennevilleetal2022} to determine the mass parameters and 3D intrinsic shape of M87. The modeling procedure used here largely follows that described in \citet{liepoldetal2023}. Our galaxy models are specified by six parameters: black hole mass \mbh, stellar mass-to-light ratio \mlgen, dark matter mass scale $M_{10}$, and three shape parameters $T$, $T_{\mathrm{maj}}$, and $T_{\mathrm{min}}$. We use a radially varying stellar mass-to-light ratio where the central \mlgen\ is a factor of 2.5 higher than at large radii {(Fig.~D1 of \citealt{liepoldetal2023})}, consistent with constraints from stellar population analyses \citep{sarzietal2018,oldhamauger2018}. The dark matter follows a generalized NFW profile with a flat inner density profile ($\gamma = 0$) and a scale radius $r_s=15$ kpc. We specify the mass scale of the dark matter halo with its enclosed mass within 10 kpc, $M_{10}$. We have made several performance improvements to \texttt{TriOS} that substantially reduced the cost of modeling. We have tested \texttt{TriOS} on simulated galaxy kinematics and demonstrated its ability to reliably recover the known input \mbh\ and other galaxy parameters in the test models \citep{pilawaetal2024}.

For each model, we compute an orbit library composed of ${\sim}350,000$ stellar orbits. The library samples 96 energies; 54 and 27 values of the second integral of motion for loop orbits and box orbits, respectively; and 27 values of the third integral of motion. The 96 energies correspond to orbits with equipotential radii logarithmically spaced from $R = 0\farcs1$ to $R = 316\arcsec$. Each orbit is projected onto the sky, convolved with the respective PSFs for the JWST and Keck datasets, and binned into the same velocity bins as the kinematic data. To ensure smoothness in the model distribution functions, we bundle groups of 27 orbits that are neighbors in phase space and average their properties before determining the orbit weights. In total, this yields 12,960 bundles of orbits. 

Orbit weights are determined via non-negative least squares \citep{lawsonhanson1995}, minimizing the $\chi^2$ over 4944 kinematic constraints (eight moments per bin; Section~\ref{sec.stellar_kinematics}). Following \citet{liepoldetal2020}, we additionally constrain $h_9$--$h_{12}$ to zero, with uncertainties matched to those of $h_7$ (odd moments) and $h_8$ (even moments), to suppress spuriously peaked LOSVDs. All kinematic moments are point-symmetrized following the scheme described in \citet{vandenboschdezeeuw2010}.

The model is required to reproduce the observed surface brightness profile and the deprojected 3D mass distribution to within 2\%. See Appendix~C of \citet{liepoldetal2023} for a description of the HST and ground-based imaging data of M87 used for this part of the modeling.

\subsection{Parameter Inference with Convex Hull and Deep Ensembles}
\label{sec.modeling.inference}

We search the 6D parameter space for the galaxy model that best reproduces the observed JWST and Keck stellar kinematics. We use a revised parameter inference scheme that improves upon the efficiency of our recent work (e.g., \citealt{pilawaetal2022, liepoldetal2023, liepoldetal2025}). The search begins with a coarse, uniform Latin hypercube sampling of ${\sim} 1000$ points spanning a broad parameter space, with each model evaluated using the TriOS code to obtain its associated $\chi^2$. We then use the convex hull algorithm to compute the 6D volume, $V_0$, of the smallest convex polytope enclosing the initial set of models. Over successive iterations, we reduce the search volume by a factor $f$, selecting at each step the $\chi^2$ threshold such that the convex hull of all models below that threshold has volume $V_i = f V_{i-1}$. Within each reduced volume, $N$ new models are drawn via Latin hypercube sampling and evaluated with \texttt{TriOS}, progressively zooming in on the best-fit region. We perform 10 iterations with $f = 0.3$ and $N = 100$, so that the final search volume is $V_{10} \sim 10^{-6} V_0$.

To sample the posterior distribution, we construct a smooth interpolation of the $\chi^2$ surface across the parameter space, then pass the corresponding log-likelihood surface to \texttt{dynesty} \citep{speagle2020} to sample, assuming a uniform prior in each parameter. To perform the interpolation, we compare three methods: Gaussian process regression \citep{rasmussenwilliams2006,pedregosaetal2011}, Student-t process regression \citep{shahetal2014}, and Deep Ensembles \citep{lakshminarayanan_simple_2017}. Gaussian process regression is a common choice for handling irregularly spaced points with noise estimations. The Student-t process regression is similar but features heavier tails, making it more robust against outliers and non-Gaussian noise in the $\chi^2$ surface. Deep Ensembles use a collection of neural networks for interpolation, adopting the variance across the ensemble as a measure of uncertainty.

\begin{deluxetable}{ll}
\tablewidth{\columnwidth}
\tablecaption{Mass and Shape Properties of M87\label{table:bf_parameters}}
\tablehead{
  \colhead{Parameter (units)} & \colhead{Inferred value}
}
\startdata
Black hole mass \mbh\ ($10^9$ $M_\odot$) & $5.38 \pm 0.13 \pm 0.22$ \\
Mass-to-distance ratio $\theta_g$ ($\mu$as) & $3.16\pm 0.08$\\
Outer \ml\ ($V$-band; $M_\odot / L_\odot$) & $3.475 \pm 0.023 \pm 0.14$ \\
Inner \ml\ ($V$-band; $M_\odot / L_\odot$) & $8.69 \pm 0.06 \pm 0.35$ \\
\parbox[t]{0.56\columnwidth}{\raggedright\strut Stellar mass within 10 kpc\newline ($10^{11}$ $M_\odot$)\strut} & $1.844\pm0.014 \pm 0.08$ \\
Total stellar mass ($10^{11}$ $M_\odot$) & $4.66\pm0.86 \pm 0.19$ \\
\parbox[t]{0.56\columnwidth}{\raggedright\strut Dark matter mass within 10 kpc\newline ($M_{10}$; $10^{11}$ $M_\odot$)\strut} & $3.57 \pm 0.10 \pm 0.15$ \\
\parbox[t]{0.56\columnwidth}{\raggedright\strut Dark matter fraction within\newline 10 kpc (\%)\strut} & $65.3 \pm 0.7$ \\
Shape parameter $T$ & $0.59 \pm 0.01$ \\
Shape parameter $T_\mathrm{maj}$ & $0.53 \pm 0.02$ \\
Shape parameter $T_\mathrm{min}$ & $0.53 \pm 0.02$ \\
\parbox[t]{0.56\columnwidth}{\raggedright\strut Average apparent-to-intrinsic\newline long axis ratio $u$\strut} & $0.935\pm 0.004$ \\
\parbox[t]{0.56\columnwidth}{\raggedright\strut Average middle-to-long\newline axis ratio $p$\strut} & $0.870 \pm 0.004$ \\
\parbox[t]{0.56\columnwidth}{\raggedright\strut Average short-to-long axis ratio $q$\strut} & $0.739\pm 0.006$ \\
Line-of-sight direction $\theta$, $\phi$ ($^\circ$) & $52.7_{-0.7}^{+1.0}$,
$35.9_{-0.9}^{+1.2}$
\\
Rotation about line of sight $\psi$ ($^\circ$) & $-66.6\pm1.1$ \\
\enddata
\tablecomments{Mass and shape parameters of M87. The first listed uncertainty is statistical, representing the 68\% credible intervals. For the mass parameters, we include a second uncertainty due to the uncertainty in the distance to M87: $16.8 \pm 0.7$ Mpc.}
\end{deluxetable}

In our tests, Deep Ensemble interpolation appears the most robust, producing less variation in the posterior distributions when its fitting hyperparameters are varied and exhibiting less sensitivity to outliers in the $\chi^2$ surface. We therefore adopt it as our fiducial interpolation scheme. The other two methods yield similar results. Across all fitting hyperparameters and methods we tested, systematic uncertainties associated with the interpolation are approximately half the size of the statistical uncertainties.

\begin{figure}
  \centering
  \includegraphics[width=\columnwidth]{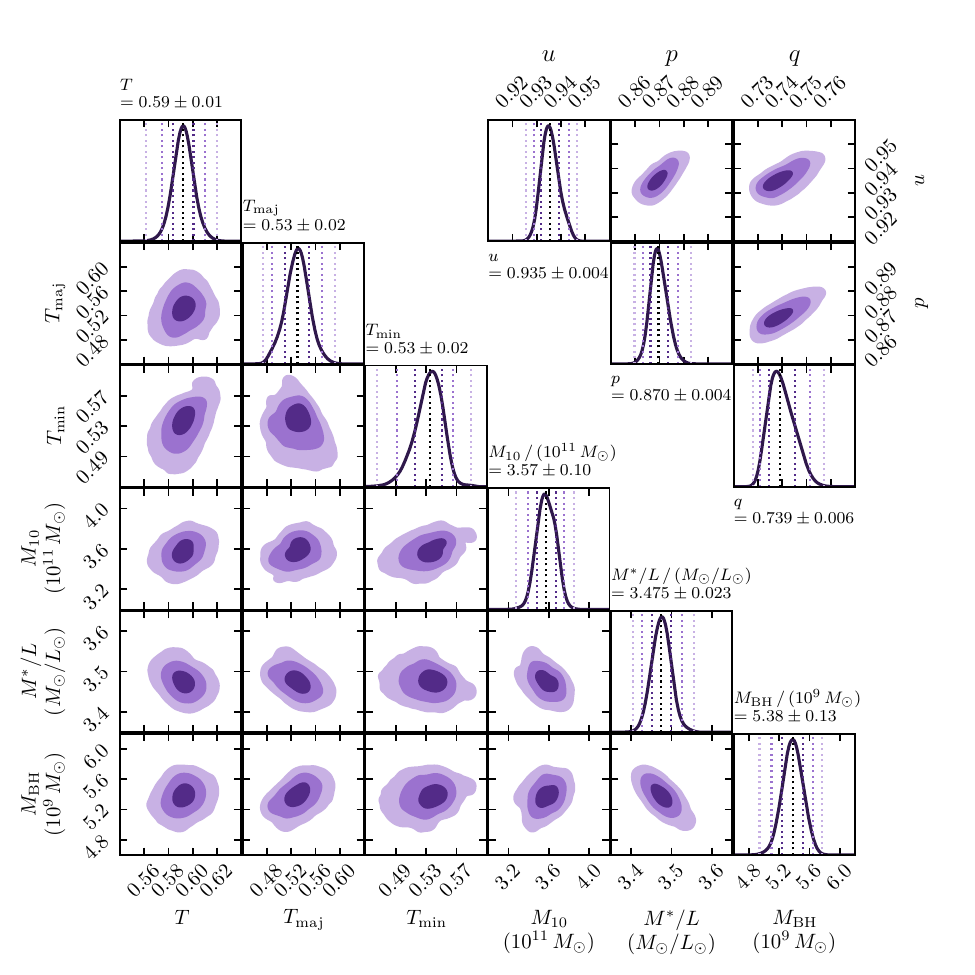}
  \caption{
    Posterior distributions of the six galaxy parameters in the triaxial orbit models of M87: black hole mass, \mbh, (outer) stellar mass-to-light ratio, \ml, dark matter halo mass enclosed within 10 kpc, $M_{10}$, and stellar halo shape parameters $T, T_{\rm maj}$, and $T_{\rm min}$.
	An alternative set of shape parameters is shown in the upper right corner: the luminosity-averaged axis ratios $p$ (middle-to-long axis), $q$ (short-to-long axis), and $u$ (apparent-to-intrinsic long axis). The three contours in each 2D panel and the vertical lines in each 1D panel mark the $68\%$, $95\%$, and $99.7\%$ credible regions.
  }
\label{fig:posterior}
\end{figure}

The final results from our triaxial orbit modeling and parameter inference procedure are listed in Table~\ref{table:bf_parameters}. The posterior distributions of the six galaxy parameters are shown in Figure~\ref{fig:posterior}. The best-fit model reproduces the observed kinematics well, including the rise in $\sigma$ and $h_4$ toward the center of M87, as plotted in Figure~\ref{fig:g_appendix}. At the posterior median, this model has $\chi^2 = 2772$ with $4944$ kinematic constraints. Naively counting the six model parameters as the number of degrees of freedom (DOF) would lead to a reduced $\chi^2$ of 0.56, but estimating the true DOF and underlying model complexity of constrained problems such as ours is a nontrivial task. The effective DOF is usually higher, as we had investigated in \citet{pilawaetal2024}. Properties of the best-fitting model are discussed in the next section.

\section{M87 Properties}
\label{sec.properties}

\subsection{Black Hole Mass, Stellar Mass, and Dark Matter Halo Mass}
\label{sec.properties.masses}

The M87 black hole mass from our combined JWST and Keck kinematics, $\mbh = (5.38 \pm 0.13 \pm 0.22)\times10^9 M_\odot$, is fully consistent with our prior result, $\mbh = (5.37^{+0.37} _{-0.25} \pm 0.22)\times10^9 M_\odot$, inferred from the Keck data alone. Notably, the statistical uncertainty in this new measurement is now below the uncertainty in the distance to M87.

The best-fit outer $V$-band stellar mass-to-light ratio, $\ml =(3.475 \pm 0.023 \pm 0.14)\, \msun/\lsun$, is within $0.4\sigma$ of our previous value. The inferred inner value is $\ml =( 8.69\pm 0.06 \pm 0.35) \, \msun /\lsun$ for the radial profile assumed in our dynamical model (see Sec.~\ref{sec.modeling.trios}). Using the V-band luminosity profile from \citet{kormendyetal2009} and our \ml\ profile, we determine the total stellar mass to be $M_* = (4.66\pm0.86\pm 0.19)\times10^{11} \, \msun$. This value is substantially smaller than was found in \citet{murphyetal2011}, $M_*{\sim}1.2\times10^{12}\msun$, which is derived from the same photometry but a much higher $\ml = 9.7 \,\msun /\lsun$ (corrected to 16.8 Mpc) obtained from axisymmetric orbit modeling. 

The gravitational sphere of influence (SOI) of a black hole is often defined as $M_*(< \rsoi)=\mbh$ or $M_*(< \rsoi)=2\mbh$. Using our mass measurements, we find M87's SMBH to have an SOI radius of $\rsoi = 4\farcs35$ ($354$ pc) or $ 6\farcs05$ ($492$ pc), comparable to M87's stellar core size of $5\farcs66-5\farcs80$ ($460-470$ pc; \citealt{laueretal2007b,dullo2019}).

The dark matter halo mass within the radial range of our kinematic data is $M_{\rm{ DM}}(<10\ {\rm kpc}) = (3.57 \pm 0.10 \pm 0.15) \times 10^{11}\ \msun$, which is  $f_{10} = (65.3 \pm 0.7) \%$ of the total mass within 10 kpc.

The \mbh\ and $M_*$ measurements reported here place M87's SMBH  above the mean \mbh\text{--}$M_*$ relation from \citet{mcconnellma2013} by 0.58 dex and outside the 0.34 dex intrinsic scatter of the relation. Agreement with core-size scaling relations is better, with M87's SMBH lying 0.15 dex and 0.12 dex below the \mbh\text{--}$r_b$ and $r_b$\text{--}$r_{\mathrm{SOI}}$ relations from \citet{thomasetal2016}, respectively, both well within the intrinsic scatter of those relations (0.29 and 0.17 dex, respectively). 

To place the M87 SMBH on the \mbh\text{--}$\sigma$ relation, we compute the luminosity-weighted average $\sigma$ within the effective radius (Equation~1 of \citealt{mcconnellma2013}), adopting $R_e \sim 80\arcsec$. Averaging over all radii down to $R = 0$, we find $\sigma_e = 304$ \kms, which drops by approximately $10$\% to 264--270 \kms\ when excluding contributions from within the SOI. M87 lies $0.39$ dex above the mean \mbh\text{--}$\sigma$ relation when including the SOI and $0.7$ dex above when excluding it, in excess of the relation's intrinsic scatter of $0.38$ dex.

\subsection{3D Shape Parameters}
\label{sec.properties.shape}

Triaxial orbit modeling enables us to obtain simultaneous constraints on the mass and shape parameters of M87. Unsurprisingly, adding the inner JWST kinematic constraints does not change the result from our prior Keck study that the stellar halo of M87 is strongly triaxial. The best-fit model in this work has a triaxial parameter of $T=0.59 \pm 0.01$, strongly ruling out an oblate axisymmetric ($T=0$) or a prolate axisymmetric ($T=1$) shape.

We note that in our MGE parameterization of M87's isophotes, all MGE components are assumed to share the same PA (i.e., no isophotal twists), but each deprojected component is allowed to have its own values of 3D intrinsic axis ratios $p=b/a$ and $q=c/a$, where $a, b$ and $c$ are the lengths of the principal axes of M87's stellar halo.\footnote{As discussed in Sec.~2 of \citet{quennevilleetal2022}, all MGE components share the same $T, T_{\rm maj}$ and $T_{\rm min}$ regardless of the values of $p$ and $q$ for each component (as long as there is no photometric twist).} We compute the luminosity-weighted averages of $p$ and $q$ across the MGE components and show their posteriors and inferred values in Figure~\ref{fig:posterior} and Table~\ref{table:bf_parameters}. The associated viewing angles are also listed in Table~\ref{table:bf_parameters}.

The axis ratios found here agree with our prior Keck study to within 3\%. While the differences exceed the formal statistical uncertainties, these uncertainties are small in part because the plausible ranges of the shape parameters are themselves narrow, given the inequality constraint of $q \le p \le u \le 1$ \citep{quennevilleetal2022}. For instance, the uncertainty on $q = 0.739 \pm 0.006$ represents only 0.8\% of its central value but 2\% of its plausible range ($0.5 \lesssim q \lesssim 0.8$).

\subsection{Orbital Structure}
\label{sec.properties.anisotropy}

\begin{figure}
  \centering
  \includegraphics[width=\columnwidth]{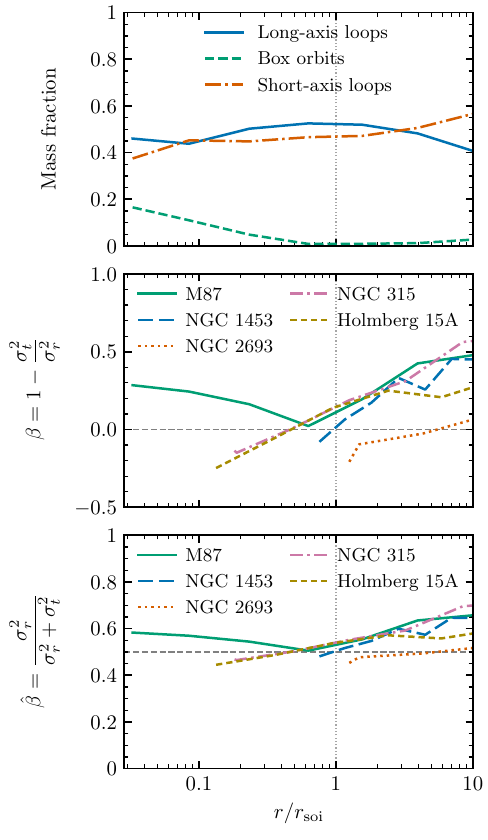}
  \caption{
    Profiles of the stellar orbit composition (top) and velocity anisotropy parameters $\beta$ (middle) and $\hat{\beta}$ (bottom) from our best-fitting models of M87.
    Radii are normalized by each galaxy's SMBH SOI radius ($r_{\mathrm{SOI}} = 6\farcs05$ for M87).
    The horizontal dashed lines indicate isotropy: $\beta=0$ and $\hat{\beta}=0.5$.
    Outside the SOI, M87's velocity anisotropy (green) becomes increasingly radial, similar to that found in triaxial orbit models of four other massive elliptical galaxies.
    Inside the SOI, M87 is somewhat radially biased.
  }
\label{fig:betahat}
\end{figure}

In the top panel of Figure~\ref{fig:betahat}, we show the profile of the orbital composition of the best-fitting orbit models. Among the three major classes of stellar orbits present in a triaxial potential, the long-axis and short-axis loop orbits have a definite sense of rotation and carry a net angular momentum along the galaxy's long (major) and short (minor) axes, respectively. The top panel shows that the stellar orbits in M87 are composed primarily of loop orbits, with very similar contributions from the short-axis and long-axis loops. Box orbits, which have zero net angular momentum, contribute only a small fraction of the total, up to ${\sim}15\%$ in the central region. 

In the lower two panels of Figure~\ref{fig:betahat}, we plot the profiles of the velocity anisotropy parameters, $\beta = 1 -\sigma_t^2/\sigma_r^2$ (middle) and $\hat{\beta} = \sigma_r^2/(\sigma_r^2+\sigma_t^2) = 1/(2-\beta)$ (bottom), of the best-fitting orbit models, where $\sigma_r$ is the velocity dispersion in the radial direction, and the tangential velocity dispersion, $\sigma_t$, is defined to be $\sigma_t^2=(\sigma_\theta^2+\sigma_\phi^2)/2$. The standard $\beta$ parameter is highly asymmetric in representing tangential vs. radial motions, with $\beta\rightarrow -\infty$ for purely tangential orbits while $\beta=+1$ for purely radial orbits. We favor the alternative measure, $\hat{\beta}$, which ranges between 0 and 1 bounded by these two limits, and is 0.5 for isotropic orbits.

At nearly all radii, the stellar motions are slightly biased radially, with $\hat{\beta}{\sim}0.6$ for both the innermost region ($\beta = 0.3$; $r\la{0.1 \rsoi}\sim0\farcs5$) and the outer region ($r\ga 3 \rsoi \sim 15\arcsec$). Near the SOI of the SMBH, the model becomes nearly isotropic. The relative sparsity of box orbits suggests that the radial anisotropy in the central region is driven primarily by loop orbits with high eccentricity, rather than box orbits. For comparison, we include the orbital anisotropy profiles for four other galaxies that we have modeled with the triaxial Schwarzschild method in Figure~\ref{fig:betahat}: NGC~1453 \citep{quennevilleetal2022}, NGC~2693 \citep{pilawaetal2022}, NGC~315 \citep{pilawaetal2025}, and Holmberg 15A \citep{liepoldetal2025}. While some of these galaxies have a mild preference for tangential orbits in the inner regions, the values of $\hat{\beta}$ differ from that of M87 only modestly, with $\hat{\beta}{\sim}0.45$ ($\beta{\sim}-0.2$) for NGC~315 and Holmberg~15A. 

Prior dynamical studies using axisymmetric orbit codes have frequently found tangential bias in the inner regions of massive cored ellipticals, interpreting it as a signature of loss-cone depletion of radial orbits by binary SMBH scouring (e.g., \citealt{thomasetal2016}). Triaxial potentials, however, support centrophilic orbit families (such as box orbits) that can continuously replenish the depleted loss cone, driving SMBH binaries toward coalescence and contributing to the stochastic gravitational wave background detected by Pulsar Timing Array experiments (e.g., \citealt{agazieetal2023}). The mild radial bias found here in M87's inner regions may reflect partial refilling of the scoured core by these centrophilic orbits, a process naturally captured by triaxial but not axisymmetric dynamical models.

The mild radial bias in the best-fit M87 model is consistent with the positive $h_4$ observed in the innermost JWST data (Figure~\ref{fig:g}). For a fixed luminosity density and potential, radially biased orbits can produce more peaked LOSVDs with extended wings than isotropic models, increasing $h_4$ relative to isotropy (e.g., \citealt{gerhard1993, vealeetal2017}). This arises because stars near the closest approach of the line of sight to the galaxy center---where the luminosity density is highest---have radial motions nearly perpendicular to the line of sight, enhancing the LOSVD peak, while stars at greater distances along the same sightline contribute radial motions at higher velocities, populating the wings.

\section{Summary and Discussion}
\label{sec.summary}

We have obtained a new measurement of M87's SMBH mass by combining our Keck KCWI stellar kinematic measurements from \citet{liepoldetal2023} with new measurements from JWST NIRSpec. The NIRSpec stellar kinematic constraints extend inward to projected radii of 10~pc, just 3\% of the SMBH's gravitational sphere of influence. Together, these data provide a contiguous high-S/N 2D map of the stellar motions over three decades in radius. Triaxial orbit models reproduce these ${\sim}$5,000 kinematic constraints well. A six-dimensional parameter search over the SMBH mass, the stellar mass-to-light ratio, the dark matter halo mass, and the intrinsic shape yields $\mbh = (5.38 \pm 0.13 \pm 0.22) \times 10^9 \msun$.

These measurements combine exquisite spatial resolution, broad spatial coverage, and high S/N. Together with our triaxial orbit models, which faithfully reproduce these observations, they provide the most comprehensive view of the stellar dynamics of M87 to date. Our analysis thus far has focused on stellar motions and not stellar populations. For the latter, we have incorporated prior stellar-population measurements in our assumed radial profile of the \ml\ parameter (see Sec.~\ref{sec.modeling.trios}). A joint study of spatially resolved stellar kinematics and population data would be an interesting next step for M87.

\begin{figure}
\centering
\includegraphics[width=\columnwidth]{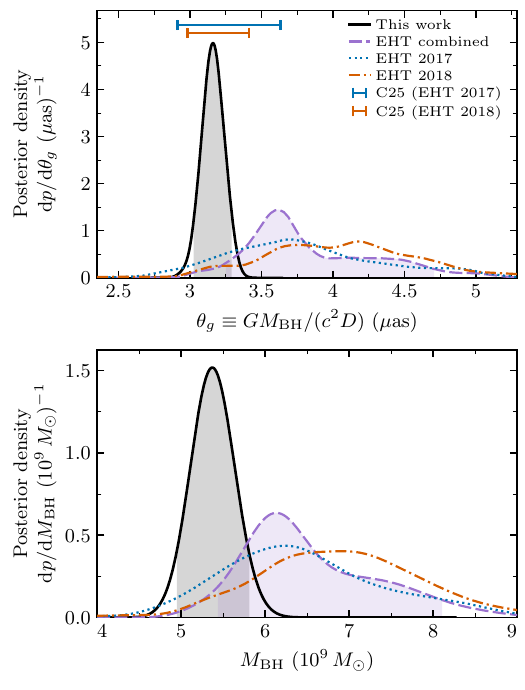}
\caption{
  Posterior distributions of M87's SMBH mass-to-distance ratio, $\theta_g \equiv G\mbh / (c^2D)$ (top), and mass, \mbh\ (bottom). Our stellar-dynamical measurement (black solid) is compared with EHT results from  2017 (blue dotted), 2018 (orange dot-dashed), and the two epochs combined (purple dashed).
  The EHT posteriors combine fits from three GRMHD pipelines (lower-right panel of Figure~9 in \citealt{EHT_M87_2025A}).
  Shaded regions mark 90\% credible intervals from the posteriors. 
  The lower panel includes the distance uncertainty $D=16.8\pm 0.7$ Mpc.
  A separate analysis of the same 2017 and 2018 EHT data by \citet{changetal2025} (`C25' in legend) yields lower $\theta_g$, as indicated by the horizontal blue and orange bars, which also indicate the 90\% credible intervals (the vertical position is arbitrary).
  }
\label{fig:moverd}
\end{figure}

The mass of the M87 black hole has been independently constrained through horizon-scale observations by EHT. In the initial analysis of their 2017 data, the EHT Collaboration measured the angular size of the gravitational radius (also referred to as the mass-to-distance ratio) from lensed emission surrounding the black hole to be $\theta_g \equiv G\mbh/(c^2 D) = 3.8 \pm 0.4$ $\mu$as. The corresponding SMBH mass is $\mbh = (6.5 \pm 0.2 \pm 0.7)\times 10^9 \msun$ \citep{EHT_M87_2019F}. In the subsequent multi-epoch analysis of the 2017 and 2018 data, the same team used an expanded library of GRMHD simulations and a newly developed Bayesian multi-epoch scoring framework, obtaining a combined mass-to-distance ratio of $\theta_g = 3.62^{+0.41}_{-0.34}$ $\mu$as, corresponding to $\mbh = 6.16^{+0.71}_{-0.57}\times 10^9\msun$ \citep{EHT_M87_2025A}. The slightly lower central value relative to 2019 was attributed in part to a preference for retrograde accretion models in the multi-epoch scoring, which tend to produce less compact images and correspondingly lower mass estimates.  An independent analysis by \citet{changetal2025}, who fit a semi-analytic dual-cone synchrotron emission model directly to the EHT visibility data using Bayesian inference, found $\theta_g$ within the 95\% highest probability density interval of $(2.84, 3.75) \mu$as for the 2017 data and $(2.93, 3.44) \mu$as for the 2018 data. Notably, while $\theta_g$ remained stable across the two epochs despite striking differences in the observed ring morphology, several other model parameters (e.g., inferred spin-axis orientation, fluid velocity direction) shifted significantly, highlighting that intrinsic variability introduces systematic uncertainties that are not yet fully captured by current models.

Figure~\ref{fig:moverd} compares the posterior of our dynamically determined $\theta_g$ (top panel) and \mbh\ (bottom panel) with the EHT results. While the peak of our posterior is lower than the peaks of the EHT 2017 and 2018 posteriors reported in \citet{EHT_M87_2025A}, their 90\% credible intervals overlap. Our results are also fully consistent with the lower $\theta_g$ inferred from a re-analysis of the same EHT data in \citet{changetal2025}. We suggest that $\theta_g$ from our stellar-dynamical work be used as a prior in new EHT analyses to sharpen inferences on M87's black hole spin and enable a more detailed characterization of the plasma conditions, magnetic field structures, and intrinsic variability of the accretion flow.

\begin{acknowledgements}
This work is based on observations made with the NASA/ESA/CSA JWST associated with program GO-2228. The data were obtained from the Mikulski Archive for Space Telescopes at the Space Telescope Science Institute (STScI). STScI is operated by the Association of Universities for Research in Astronomy, Inc., under NASA contract NAS 5-03127 for JWST. 
Other spectroscopic data presented in this paper were obtained at the W. M. Keck Observatory, which is operated as a scientific partnership among the California Institute of Technology, the University of California and the National Aeronautics and Space Administration. The Observatory was made possible by the generous financial support of the W. M. Keck Foundation. 
This work used Expanse at the San Diego Supercomputing Center through allocation AST180041 from the Advanced Cyberinfrastructure Coordination Ecosystem: Services \& Support (ACCESS) program, which is supported by U.S. National Science Foundation grants \#2138259, \#2138286, \#2138307, \#2137603, and \#2138296.
E.R.L. and C.-P.M. acknowledge the support of NSF AST-2206307 and the Heising-Simons Foundation. J.L.W. acknowledges the support of NSF AST-2206219. \\[2ex] 
\end{acknowledgements}

\facilities{Keck (KCWI), JWST (NIRSpec)}

\software{
\texttt{TriOS} \citep{quennevilleetal2021,quennevilleetal2022},
\texttt{dynesty} \citep{speagle2020},
\texttt{pPXF} \citep{cappellari_full_2023},
\texttt{vorbin} \citep{CappellariCopin2003},
KCWI Data Reduction Pipeline \citep{Morrisseyetal2018},
JWST Science Calibration Pipeline \citep{bushouseetal2023},
astropy \citep{astropy:2013,astropy:2018,astropy:2022},
matplotlib \citep{Hunter:2007},
numpy \citep{harris2020array},
scipy \citep{scipy},
}

\restartappendixnumbering
\appendix

\section{Correcting ``wiggle'' artifacts with \texttt{hyphen}}
\label{appendix:hyphen}

\subsection{`Wiggles' in the detector image}

The NIRSpec PSF is undersampled in the IFU and MSA modes, producing substantial resampling artifacts. The peak flux of a point source is highly dependent on its position in the pixel grid. Near the center of a pixel, its light will land primarily within that pixel, while near the edge of a pixel, the light is split across neighboring pixels, reducing the peak flux \citep{lawetal2023}. 

\begin{figure*}
  \centering
  \includegraphics{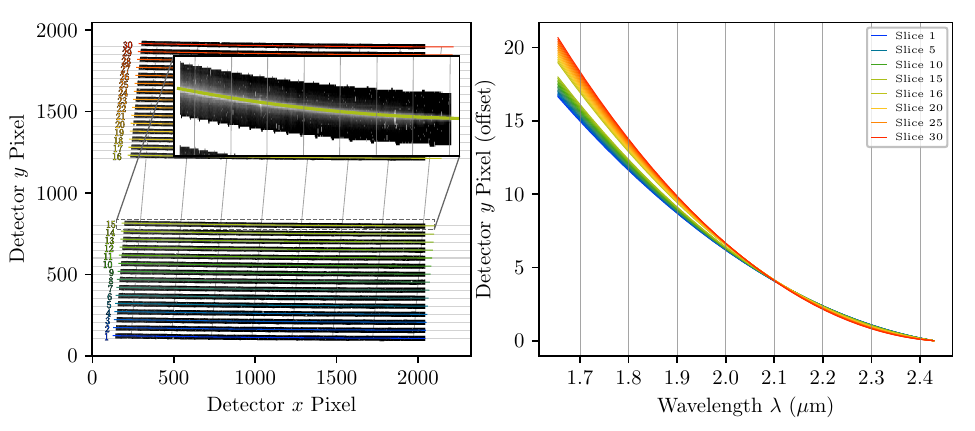}
  \caption{
    (Left) A representative NRS1 detector image.
    Colored curves mark one spectral trace in each slice.
    Lines of constant wavelength (0.1\micron\ steps) are shown as slanted grey lines.
    The spectral traces are not perfectly horizontal and are slightly curved (inset).
    (Right) The shapes of the spectral traces vary across the detector, with traces on one side crossing roughly five more detector rows than those on the other side, producing a spatially varying wiggle frequency.
 }
\label{fig:a}
\end{figure*}

On the JWST NIRSpec detector, light from a fixed sky location is dispersed approximately along the detector $x$-axis. Importantly, the spectral trace produced by the dispersed light is not perfectly parallel to the $x$-axis and is slightly curved. This is visible in the detector image in Figure~\ref{fig:a}. As these traces cross successive detector rows, the point source's light alternates between being concentrated within one row and being split between adjacent rows as the wavelength changes.

Figure~\ref{fig:a} shows the shapes of spectral traces which lie at different locations on the NRS1 detector. Traces in low-numbered slices pass through fewer detector rows than high-numbered slices, and there is a discontinuity in the trace shape between slices 15 and 16.

\begin{figure}
  \centering
  \includegraphics[width=\columnwidth]{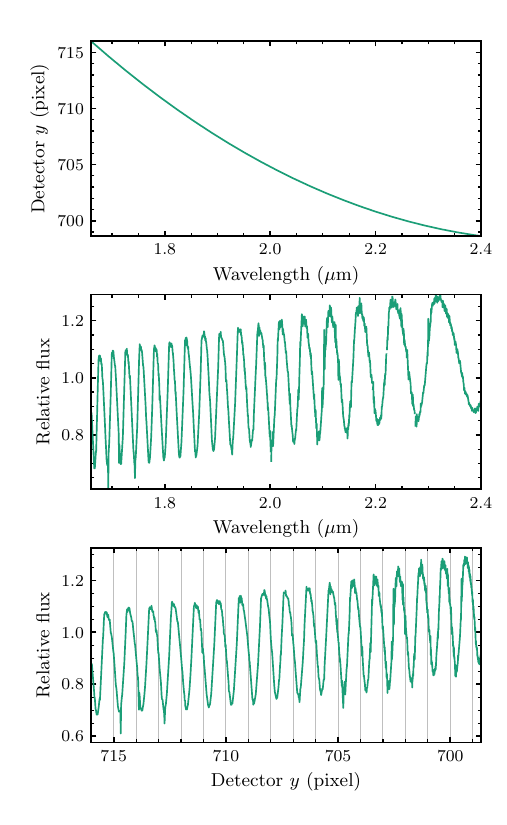}
  \caption{
    (Top) The spectral trace for a representative spaxel near the AGN on the NRS1 detector is described by $y(\lambda)$, the detector row coordinate as a function of wavelength.
    (Middle) When the spectrum is indexed by wavelength, the wiggle artifacts have a varying period, which becomes shorter near the blue side of the NRS1 detector.
    (Bottom) When the spectrum is indexed by the detector row coordinate $y$, the period of the wiggle artifacts is approximately one pixel. To guide the eye, vertical grey lines are drawn at integer values of the detector pixel index. The peaks and troughs of the wiggle pattern appear at a consistent location between each pair of grey lines. 
  }
\label{fig:f}
\end{figure}

\begin{figure}
  \centering
  \includegraphics[width=\columnwidth]{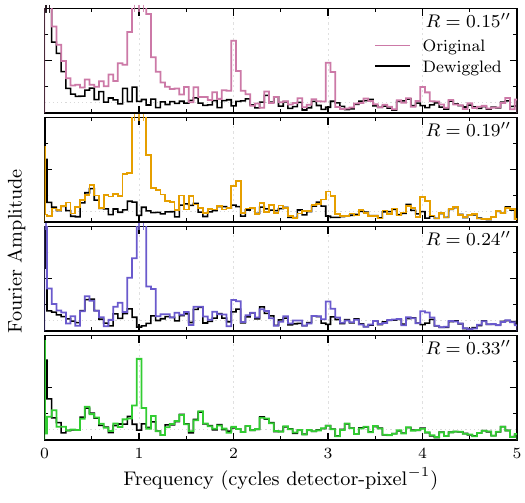}
  \caption{
    When the spectrum is indexed by the detector row coordinate $y$, the wiggles have a fundamental frequency of roughly one cycle per detector pixel. Fourier amplitudes of four wiggle-impacted spectra are shown before (colored) and after (black) hyphenation. The four spectra here are shown in Figure~\ref{fig:b}. The wiggle artifact is clearly seen as a peak centered at $f=1$ with diminishing harmonics at $f=2,3,4$. After hyphenation, these peaks are no longer present, leaving amplitudes similar to those at neighboring frequencies.
  }
\label{fig:c}
\end{figure}

The wavelength-dependent frequency of the `wiggle' artifacts has been a challenge for previous attempts to correct the effect \citep[e.g.][]{pernaetal2023, dumont_wiggle_2025, shajib_textttraccoon_2025}. The wiggle pattern originates in the spectral trace repeatedly crossing into different detector rows at different wavelengths. When the shape of the spectral trace on the detector (that is, the functions $x(\lambda)$ and $y(\lambda)$) is known, the spectra can be indexed by the detector row coordinate $y$ rather than $\lambda$. In this coordinate, the wiggles have a fixed period of approximately one pixel. Accordingly, the local frequency (in wavelength) of the wiggle pattern is $|dy/d\lambda|$. Figure~\ref{fig:f} shows a representative central spectrum indexed by wavelength and by the detector row coordinate $y$. The Fourier amplitudes show a peak at approximately one cycle per detector pixel and its harmonics, which are absent after use of our wiggle correction code \texttt{hyphen} (Figure~\ref{fig:c}).

\begin{figure*}
  \centering
\includegraphics[width=\textwidth]{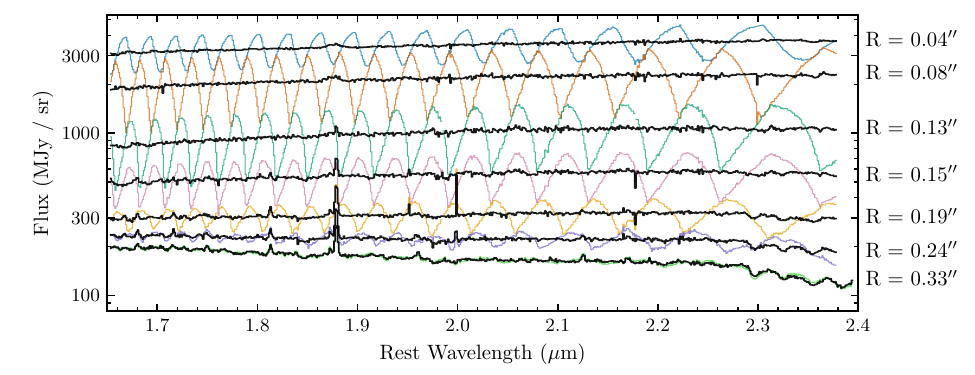}
\includegraphics[width=\textwidth]{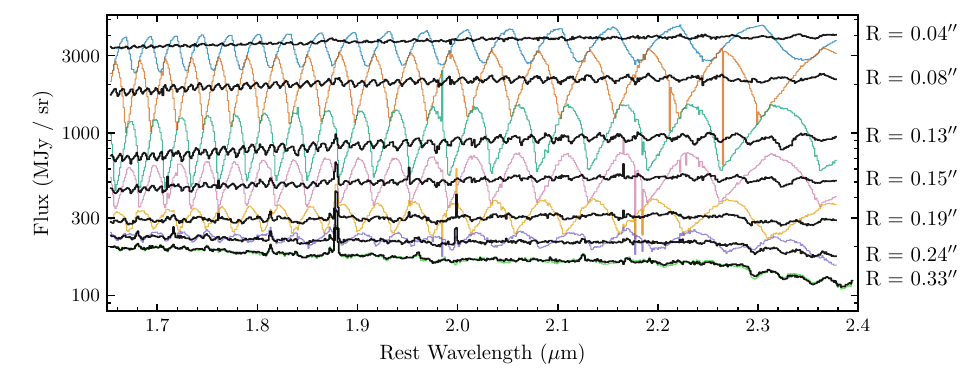}
 \caption{Comparison of uncorrected (colored) and dewiggled (black) spectra at various distances from the center of the datacube for a single NIRSpec IFU exposure. These spaxels trace a path to the NE of the center. The uppermost spectra are dominated by AGN light, making stellar kinematic measurements intractable, even after correction. CO-bandhead features become visible between 0\farcs1 and 0\farcs2 and grow more prominent at larger radii. (Top) After correction with \texttt{hyphen}, no discernible wiggle artifacts remain. (Bottom) Same as top, but now comparing uncorrected (colored) and ATM-dewiggled (black) spectra. For our data, the adaptive trace method \citep{lawclarke2026} leaves substantial residuals, most of which are at a higher frequency than the original wiggle artifacts.}
\label{fig:b}
\end{figure*}

Recently, the adaptive trace method \citep{lawclarke2026} was integrated into the JWST Science Calibration Pipeline. When applying that method to our M87 data, we found substantial residual wiggle artifacts, leaving the central part of the data unusable for the work in this paper (Figure~\ref{fig:b}). For observations of other ETGs, ATM performs similarly to our code \texttt{hyphen}. We anticipate that ATM will be sufficient in many cases, but provide \texttt{hyphen} to the community for cases where ATM leaves residual artifacts.

\subsection{The \texttt{hyphen} model}
\label{appendix.hyphen.model}

The \texttt{hyphen} model uses each exposure's 2D detector image and corresponding rectified 3D datacube. Within each cube, we identify the bright point source associated with wiggle artifacts and a target region around it (the central 0\farcs45 for our M87 analysis). WCS transformations between the image and datacube associate each target spaxel with a spectral trace on the detector, whose shape describes how the wiggle frequency varies with wavelength. The model described below fits and subtracts the quasiperiodic wiggle signal from each target spaxel's spectrum. The result is a corrected 3D datacube for each exposure. If the input cubes share a common spatial and spectral grid, the corrected cubes can optionally be combined into a single datacube. The implementation is described in Appendix~\ref{appendix.hyphen.implementation}.

In the \texttt{hyphen} code, we model the spectrum in a target spaxel $i$ as the sum of a field spectrum $\mathcal{F}_i$ and a contribution from the PSF wings of the central point source $\mathcal{P}_i$. That point-source-like spectrum is modulated by a `wiggle' term $\mathcal{W}_i$, which is quasiperiodic in wavelength with a varying frequency. When using \texttt{hyphen} to correct the spectra\footnote{As a shorthand, we refer to application of the \texttt{hyphen} method as `hyphenation' of the data. We also interchangeably refer to the \texttt{hyphen} model and code as `\texttt{hyphen}'.}, we subtract the wiggle term from the observed spectrum. That is, the observed spectrum $\mathcal{O}_i$ is modeled as $\mathcal{O}_i \approx \mathcal{F}_i + \mathcal{P}_i + \mathcal{W}_i$, and the corrected spectrum is $\mathcal{O}'_i = \mathcal{O}_i - \mathcal{W}_i \approx \mathcal{F}_i + \mathcal{P}_i$.

For each exposure, we construct a pair of reference spectra $\mathcal{F}$ and $\mathcal{P}$ by coadding spectra from a region far from the central point source (typically $1\farcs0 \leq R \leq 1\farcs5$) and a region dominated by the central point source ($R \leq 0\farcs2$), respectively. Both reference spectra are normalized to have unit median flux. The contribution of these components to the spectra of individual spaxels is then specified by a pair of fit parameters with $\mathcal{P}_i = \alpha_i \mathcal{P}$ and $\mathcal{F}_i = \gamma_i \mathcal{F}$. We typically treat $\alpha_i$ and $\gamma_i$ as constants in wavelength. In some cases where there is a mismatch between the reference spectra and the spaxel spectra, we find that the fit is substantially improved by instead modeling $\alpha_i$ and $\gamma_i$ as low-order Legendre series in wavelength. Constant envelope functions are equivalent to order-zero Legendre series. The Legendre series orders are chosen adaptively, as described in Appendix~\ref{appendix.hyphen.implementation}.

As shown in Figure~\ref{fig:f}, while the wiggle artifacts are quasiperiodic in $\lambda$, they are periodic in the detector row coordinate $y$. We model this periodic shape with a sinusoidal series:
$$w(y) = \sum_{n=1}^{N_w} \left[ a_n \sin(2 \pi n f y) + b_n \cos(2 \pi n f y)\right].$$
We allow for a frequency adjustment term $f\approx 1$ to accommodate a slight mismatch between the expected and actual base wiggle frequency. The wiggle order $N_w$ is chosen adaptively (Appendix~\ref{appendix.hyphen.implementation}). The amplitude and phase of each Fourier mode (set by the $a_n$ and $b_n$ coefficients) are left as fit parameters.

We generally allow for the wiggle amplitude to be a weak function of wavelength with a third envelope term $\beta_i \approx 1$. As with $\alpha_i$ and $\gamma_i$, $\beta_i$ is modeled as a low-order Legendre series in wavelength with an adaptively chosen order, starting with order 0. The zeroth-order term of this Legendre series is fixed to unity, since the Fourier coefficients already set the overall wiggle amplitude. The wiggle term is then $\mathcal{W}_i = w_i \beta_i \mathcal{P}_i$.

The full \texttt{hyphen} model is $$\mathcal{O}_i \approx \gamma_i \mathcal{F} + (1 + w_i \beta_i) \alpha_i \mathcal{P},$$ and the hyphenation process is $$\mathcal{O}'_i = \mathcal{O}_i - w_i \beta_i \alpha_i \mathcal{P}.$$ We denote the orders of the Legendre series $\alpha_i$, $\beta_i$, and $\gamma_i$ as $N_\alpha$, $N_\beta$, and $N_\gamma$, respectively. An order-$N$ series has $N+1$ coefficients, and the zeroth-order coefficient is the constant term of the series. In total, the \texttt{hyphen} model has $2 N_w$ free parameters from the phase and amplitude of each order of the $w_i$ series, $N_\alpha + 1$ from the $\alpha_i$ series, $N_\beta$ from the $\beta_i$ series (as the zeroth-order coefficient is fixed to 1), $N_\gamma + 1$ from $\gamma_i$, and optionally an additional parameter if $f$ is allowed to vary. This totals $2 N_w + N_\alpha + N_\beta + N_\gamma + 2$ free parameters (plus one if $f$ is allowed to vary), which must be determined from each spaxel's spectrum.

\subsection{Implementation of the \texttt{hyphen} Model}
\label{appendix.hyphen.implementation}

The \texttt{hyphen} model is implemented in efficient and flexible Python code. The code is run through a command-line interface and all configuration is specified through human-readable \texttt{yaml} text files. The code and its dependencies can be installed using the pip package installer. Documentation of the \texttt{hyphen} code is provided at the project's GitHub webpage.\footnote{\url{https://github.com/EmilyLiepold/hyphen}} We provide a brief overview of the procedure followed by the code here.

The \texttt{hyphen} code is designed to operate on individual exposures of NIRSpec IFU or MSA data, expecting both the 2D detector image and the corresponding rectified 3D datacube for each exposure. If those exposures are to be combined, then during rectification a common spatial and spectral grid should be used. Each exposure is dewiggled separately; the corrected datacubes are optionally merged into a combined datacube. 

After loading the data for a given exposure, the code first computes and normalizes the point-source and field reference spectra. The location of the point source is either provided by the user or determined by finding the brightest spaxel in the cube. Spaxels within a specified radius of the point source (typically $R \leq 0\farcs5$) are selected and their spectra from the 3D datacube are fit in parallel, as described in Appendix~\ref{appendix.hyphen.model}. For each spaxel, we use the WCS transformations between the 2D detector image and 3D datacube to determine the spectral trace at its location within the field of view. The spectral trace gives the detector coordinates $x(\lambda)$ and $y(\lambda)$ of the dispersed light associated with that 3D spaxel. We use this trace to index the spectrum by the detector row coordinate $y$.

The initial fit for each spaxel uses $N_w = N_\alpha = N_\beta = N_\gamma = 0$, with no wiggle terms and constant envelope terms. The fit is done using the trust region reflective least-squares algorithm implemented in \texttt{scipy.optimize}. We compute the root-mean-square fractional residual over the fitted wavelengths, normalizing each residual by the observed flux. If this value exceeds a user-specified threshold (typically 5\%), we increase the model complexity by adding wiggle terms or increasing one of the orders of the Legendre series. If the more complex model does not improve the fit beyond some chosen threshold (typically an absolute reduction in the RMS fractional residual of 0.01\%), we revert to the previous model and try increasing a different component. This procedure continues until the RMS fractional residual falls below the specified threshold, no increase in the series orders sufficiently improves the fit, or the maximum allowed value for each series order is reached. The final model is then used to compute the wiggle-free spectrum for that spaxel.

For each input exposure, we apply this procedure to all selected spaxels, resulting in a set of wiggle-free 3D datacubes. The code can then optionally combine these datacubes on the same spatial and spectral grid through a weighted average of the exposure fluxes, using the interpolation weights determined during rectification.  Merging uncorrected datacubes with \texttt{hyphen} reproduces the combined datacube produced by the JWST pipeline's \texttt{cube\_build} step.

\section{Stellar Kinematics of the best-fitting orbit model}\label{appendix:bfmodel}

\begin{figure*}
  \centering
\includegraphics[width=\textwidth]{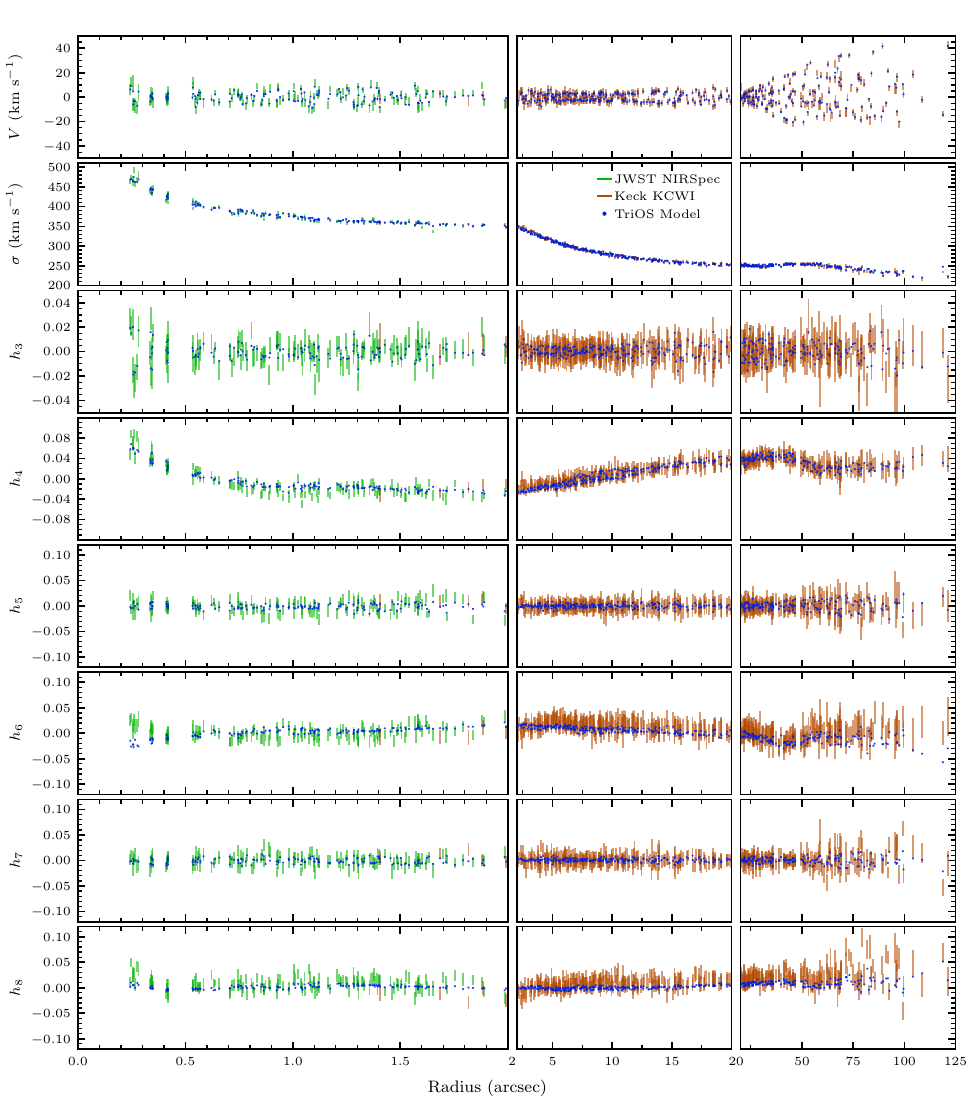}
 \caption{
  Same as Figure~\ref{fig:g}, except the point-symmetrized versions of the NIRSpec (green) and KCWI (orange) stellar kinematics are shown. These are used as constraints in the TriOS models and are well reproduced by the best-fit orbit model (blue).
 }
\label{fig:g_appendix}
\end{figure*}

The stellar kinematics of the best-fitting orbit model, as described in Section~\ref{sec.modeling}, are shown in Figure~\ref{fig:g_appendix}.

\clearpage

\bibliographystyle{yahapj}
\bibliography{M87}

\end{document}